\documentstyle[12pt]{article}

\begin{document}

\topmargin 0pt \oddsidemargin=-0.4truecm \evensidemargin=-0.4truecm 
\baselineskip=24pt 
\setcounter{page}{1} 
\begin{titlepage}     
\begin{flushright}
March 1996
\end{flushright}
\vspace*{0.4cm}
\begin{center}
{\LARGE\bf 
Radiative Seesaw Mechanism at Weak Scale} 
\vspace{0.8cm}

{\large\bf  Zhijian Tao}

\vspace*{0.8cm}

{\em Theory Division, Institute of High Energy Physics, Academia Sinica\\
\vspace*{-0.1cm}
Beijing 100039, China\\}
\end{center}
\vspace{.2truecm}     

\begin{abstract}
We investigate an alternative seesaw mechanism for neutrino mass 
generation. Neutrino mass is generated at loop level but the 
basic concept of usual seesaw mechanism is kept. One simple model 
is constructed to show how this mechanism is realized. The applications  
of this seesaw mechanism at weak scale 
to cosmology and  neutrino physics are discussed. 
\end{abstract}
\vspace{2cm}
\centerline{} 
\vspace{.3cm}
\end{titlepage}
\renewcommand{\thefootnote}{\arabic{footnote}} \setcounter{footnote}{0} 
\newpage
Seesaw mechanism \cite{yan} is one of the best and simplest ways to
understand why neutrino, if massive, is much lighter than the corresponding
charged lepton in the same generation. The central idea of the seesaw is to
introduce a right-handed neutrino $\nu _R$, which will couple to lepton
doublet through Yukawa coupling. The point is that besides the Yukawa
interaction term there is another bare Majorana mass term $M_R$ for $\nu _R$%
. After the gauge symmetry breaking the Yukawa term will result in a Dirac
neutrino mass $m_D$. Therefore the neutrino mass matrix takes the form 
\begin{equation}
\left( 
\begin{array}{cc}
0 & m_D \\ 
m_D^{+} & M_R 
\end{array}
\right) 
\end{equation}
In the three generation model $m_D$ and $M_R$ are three by three mass
matrix. Diagonalizing the mass matrix one gets the neutrino mass
eigenstates. If $M_R$ is much bigger than $m_D$ the mass of the light
neutrinos, which are mostly left-handed, is determined as $m_D^TM_R^{-1}m_D$%
. The heavy states, which are mostly right-handed, have mass almost as $M_R$%
. Therefore one sees that even if the Dirac mass term is comparable to the
charged lepton mass the light neutrino mass can be much smaller. The
features one should notice in this mechanism are the following: $M_R$ is a
free scale usually taken from weak scale to the Grand Unification Scale
(GUT). And the heavy neutrinos are not stable, they decay through mixing to
the light neutrinos. For large $M_R$ the heavy neutrinos decay very fast, so
they have no cosmological consequence. In this mechanism the lepton number
symmetry is broken either explicitly or spontaneously. Although smallness of
the neutrino mass can be understood in this mechanism, the actual values of
neutrino mass and mixing are not predicted due to the unknown scale $M_R$
and structure of $m_D$. As an indication, if one assumes that $m_D$ is same
as the charged lepton mass matrix and $M_R$ is a unit matrix up to a scale,
one gets the relations for the light neutrino masses $m_{\nu _i}=m_i^2/M_R$,
where the index i denotes the i-th generation. So it is the scale $M_R$
determines the order of the magnitude of the neutrino mass. If $M_R$ is at
GUT scale, one obtains $m_{\nu _e}<<m_{\nu _\mu }<<m_{\nu _\tau }\leq
10^{-3} $eV. These tiny masses may only play a role for solar neutrino
behavior. Another most interesting scale is the weak scale. There are a
number of physical motivation to consider $M_R$ at weak scale. First of all
for weak scale $M_R$ the new physics mechanism can be tested in the future
experiments, secondly it avoids to introduce an intermediate scale between
weak and GUT scales. For $M_R$ at weak scale all three light neutrino masses
are close to their upper bound, i.e. a few eV, 100KeV and 10MeV for
electron, muon and tau neutrinos. These neutrinos are strongly constrained
from cosmological and astrophysical consideration depending on their decay
modes \cite{gyu}. Obviously they offer no solutions to the solar neutrino
and atmospheric neutrino problems \cite{smi}, but they may play a role in
the dark matter issue by providing either a hot dark matter component in the
mixed dark matter model \cite{mdm} or a late decaying particle \cite{dod,kim}
in the cold dark matter model \cite{cdm}. And no cold dark matter candidate
is provided. Moreover there may be a problem for the seesaw model in
consideration of the baryogenesis of the universe. The problem is due to
B-L(baryon number minus lepton number) symmetry violation. Once the B-L
violation process through seesaw mechanism and the anomalous B+L process
induced by gauge interaction are in the thermal equilibrium at an early
stage of the universe \cite{kuz}, any primordial origin of baryon and lepton
asymmetries generated earlier are washed out. It leads to very strong
constrains on the neutrino mass \cite{fuk}. An upper bound of a few eV for
all three light neutrino masses are obtained in order to avoid this problem 
\cite{pec}, which in turn implies the scale of $M_R$ should be much larger
than weak scale. However, this problem can be evaded if the B-L symmetry is
spontaneously broken at the weak scale. Before B-L symmetry breaking only
B+L violating process due to gauge anomaly is active and after the B-L
symmetry breaking the anomalous B+L violating process is already suppressed
as the temperature of universe is low enough. These two processes will never
be in thermal equilibrium through the evolution of the universe. Hence the
constrains on the strength of B-L violation from the baryogenesis of the
universe is avoided \cite{tur}.

In this letter we consider a different scheme for seesaw mechanism. The main
consideration is to keep the basic concept of the seesaw mechanism, i.e. the
light neutrino mass is suppressed by the large right-handed neutrino mass $%
M_R$, and require the neutrino mass only generated radiatively 
\cite{bau}. For this
kind of scenario neutrino mass is expressed as $m_\nu \sim \frac \lambda
{16\pi ^2}m_D^TM_R^{-1}m_D$. One sees that adding to the usual seesaw form
is another suppression factor from loop effect. $\lambda $ is some
combination of the coupling constants besides Yukawa coupling. This constant
can be very small naturally if it is associated with the lepton number
violation. Therefore the neutrino mass is at least two order of magnitude
smaller than that in the usual seesaw model for the same scale $M_R$. This
scenario has some very interesting features. First of all $\nu _R$ can be
stable by imposing some discrete symmetries, while still giving light
neutrino nonzero mass. In fact, in order to avoid tree-level Dirac neutrino
mass these symmetries are necessary. This is very different from the
original seesaw mechanism, where $\nu _R$ is unstable for nonzero light
neutrino mass. The application of the stable $\nu _R$ is to play a role of
the cold dark matter. Secondly light neutrino mass is suppressed also by the
loop effect, so for a weak scale $M_R $ the neutrino mass can be much
smaller than the current experimental bounds. The light neutrino may be
provided as a candidate for explaining the solar neutrino, atmospheric
neutrino problems and a hot dark matter component in the mixed dark matter
model or still a late decaying particle in the cold dark matter model.
Baryogenesis of the universe also restricts this kind of mechanism, but the
constrains are relaxed due to the loop factor. The most attractive picture
is that if there is no other scale except weak scale below GUT, $M_R$ is
around this scale, then in this scenario with only one scale and with only
neutrino particles, one may explain the observed dark matter problem and the
structure formation of the universe, and possibly other related phenomena in
neutrino physics. From now on we will call the original seesaw mechanism as
tree-level seesaw mechanism and the other radiative seesaw mechanism.

Now let us implement this mechanism in a very simple model. This model is to
extend the standard model by introducing three family of right-handed
neutrinos $\nu _R$ and one more Higgs doublet $\Phi $. We impose an $Z_2$
discrete symmetry on this model. Under this symmetry transformation $\nu _R$
and $\Phi $ change sign, while other fields remain same. As a result of this
symmetry, $\nu _R$ does not couple to the standard Higgs $\Phi _S$ through
Yukawa coupling. Only the new Higgs doublet $\Phi $ couples to $\nu _R$. Now
we can write down all the possible interaction terms for this model. It
includes the gauge interaction, Yukawa coupling and the Higgs potential.
However in this work only the Yukawa interaction for lepton and part of the
Higgs potential are relevant. The Yukawa coupling is expressed as following: 
\begin{equation}
L_Y=f_{ij}\bar l_ie_{Rj}\Phi _S+g_{ij}\bar l_i\nu _{Rj}\Phi +h.c+M_{ij}\nu
_{Ri}^T\nu _{Rj} 
\end{equation}
Here $l_i$ and $e_{Ri}$ are lepton doublet and the right-handed charged
lepton respectively. Since the $Z_2$ symmetry is exact and will not be
broken, $\Phi $ will not develop a nonzero vacuum expectation value (VEV).
Therefore with only $L_Y$ lepton number is not broken, i.e. neutrino does
not obtain mass at this level. However with all the terms in $L_Y$ and a
term like $\lambda (\Phi _S^{+}\Phi )^2$ in the potential, it is easy to
check that lepton number symmetry is not automatically conserved anymore. In
other words neutrino must develop a nonzero mass, but obviously this mass is
generated only at loop level. If the masses of $\Phi $ and $\nu _R$ are at
the same order of the magnitude $M_R$, the light neutrino mass can be
estimated as, up to a logarithmic factor, 
\begin{equation}
m_\nu \simeq \frac \lambda {16\pi ^2}g^TM_R^{-1}gV^2 
\end{equation}
where $V$ is the VEV of the standard Higgs $\Phi_S $. In the tree-level
seesaw mechanism it is assumed that the couplings $g$ and $f$ have same
order of magnitude and similar structure, that is $gV\sim fV=m_D$. Then in
this model the basic seesaw concept is realized at the loop level. Compared
with the simplest tree level seesaw model, which is the standard model plus
right-handed neutrino, our model only has one more Higgs doublet and
introduce an additional $Z_2$ discrete symmetry. Because the $Z_2$ symmetry
is not broken, $\Phi _S$ and $\Phi $, $\nu _R$ and $\nu _l$ will not mix
with each other respectively. The lightest particles among $\nu _R$ and $%
\Phi $ are stable. In above description the lepton number is explicitly
broken. Of course the lepton number may also be broken spontaneously with
introduction of some singlet scalar fields as in the singlet majoron model 
\cite{pec1}. In the singlet majoron model the mass term of $\nu_R$ is
replaced by $h\nu_R\nu_RS$, where $S$ is the singlet scalar field. When $S$
field gets a nonzero VEV, lepton number is spontaneously broken. The
difference between tree-level seesaw model and our model is that in our
model the majoron only couples to $\nu _R$ at tree level. The light neutrino
couples to majoron not through mixing but radiative correction.

Now we come to discuss the application of our model to the dark matter issue
of the universe and other issues in neutrino physics. The very interesting
question is to see how the right-handed neutrino may serve as the candidate
of cold dark matter. In our model in principle both $\nu _R$ and $\Phi $ can
be the candidate of the cold dark matter depending on which particle is the
lightest one. Here we assume that one of $\nu _R$ is the lightest particle
among $\nu_R$ and $\Phi$, and from now on we just call it $\nu_R$. Because $%
\Phi $ has direct standard gauge coupling to Z boson, if it is the dark
matter the elastic scattering cross section of $\Phi $ from the nuclei of
the detector is determined by this neutral current interaction. Having not
observed any signal of this reaction requires the mass of $\Phi $ to be at
least a few TeV \cite{cal}. On the other hand $\nu _R$ dark matter is not
constrained much from the direct dark matter search experiments. The relic
abundance of $\nu _R$ is controlled by its interaction with other light
particles and the evolution of the universe. We are going to estimate the
relic density of $\nu _R$ in our model with and without majoron.

First let us see the case without majoron. Most generally the evolution of $%
\nu _R$ is determined by the combined evolution equations of $\nu _R$ and $%
\Phi $. The equations include the contributions from $\bar \nu _R\nu _R$
annihilation, $\Phi \Phi $ annihilation and the decay from $\Phi $ to $\nu _R
$. If $\Phi $ is not almost degenerate with $\nu _R$, i.e. the mass
difference $\Delta M$ is significantly larger than the freezeout temperature
of $\nu _R$, one may neglect the presence of $\Phi $. Then only the $\bar
\nu _R\nu _R$ annihilation cross section $<\sigma _Av>$ determines the relic
density of $\nu _R$. Approximately the contribution of the thermal relics of
a massive cold dark matter particle to mass density of the universe can be
expressed as $\Omega h^2\sim 10^{-37}cm^2/<\sigma _Av>$ \cite{kol}. To be
the candidate of the cold dark matter, its annihilation cross section should
be roughly as large as $10^{-37}cm^2$. And the freezeout temperature $%
T_D(\nu _R)$ at which $\nu _R$ decouples from the thermal equilibrium is
about $M_R/20$. For ${\bar \nu _R}\nu _R$ annihilation the dominant channel
is ${\bar \nu _R}\nu _R\to {\bar \nu _l}\nu _l,~~e^{+}e^{-}...$, which is
related to the light neutrino mass generation. The cross section for this
channel is estimated as 
\begin{equation}
<\sigma _Av>\sim \frac{\lambda ^{\prime }{}^2m_D^4}{\pi M_R^6}=10^{-39}(
\frac{\lambda ^{\prime }}{1.0})^2(\frac{m_D}{1.7GeV})^4(\frac{100GeV}{M_R}%
)^6cm^2
\end{equation}
where $\lambda ^{\prime }$ represents all possible contribution from the
scalar potential including $\lambda $ term. Since $\lambda $ is related to
the lepton number violation, $\lambda ^{\prime }$ can be naturally much
larger than $\lambda $. In fact with the parameters chosen reasonably as in
above equation, the annihilation cross section is much smaller than what
needed for $\nu _R$ being the cold dark matter, $\nu _R$ contribution
overcloses the universe. On the other hand, however, if $\Delta M$ is much
smaller than $T_D(\nu _R)$, the density of $\Phi $ and $\nu _R$ are both
determined by annihilation process $\Phi \Phi \to $ light standard model
particles. The cross section is estimated as 
\begin{equation}
<\sigma v>\simeq \frac{\pi \alpha ^2}{M_R^2}\simeq 10^{-35}(\frac{100GeV}{%
M_R})^2cm^2
\end{equation}
where $\alpha $ is the fine structure constant. At this extreme situation
with the parameters chosen as in the equation (5), 
the annihilation process is
too strong, it contributes only a small portion of needed dark matter
density. Although for $\Delta M$ between these two extreme situation 
one needs to
solve the combined evolution equations, 
one can certainly expects for a certain range
of $\Delta M$ from $T_D(\nu _R)$ to $M_R$, the relic $\nu _R$ is able to
contribute a closure density to the universe. A similar case is
investigated quantitatively but in a different model \cite{cha}. 
Its numerical
calculation supports this expectation in our model.

As we already mentioned the light neutrino mass depends on the parameter $%
\lambda$. With $\lambda\leq 1$, one obtains $m_{\nu_{\tau}}\leq 100$ KeV, $%
m_{\nu_{\mu}}\leq 300$ eV, $m_{\nu_{e}}\leq 10^{-2}$ eV. We investigate
three possible choices for neutrino mass, which are interesting in neutrino
physics. The first is $m_{\nu_{\tau}}\simeq 5$eV, then $\nu_{\tau}$ can be
the hot dark matter component needed for the large scale structure formation
in mixed cold dark matter model. In this case $\nu_{\mu}$ mass is close to $%
10^{-2}$eV and $\nu_e$ is very light as expected from seesaw. If the mass of 
$\nu_{\mu}$ is a few times smaller than $10^{-2}$eV, $\nu_{\mu}$, $\nu_e$
oscillation may offer a solution to solar neutrino problem through MSW
mechanism. If it is a few times larger, the mass square difference for these
two neutrino species are just what needed for atmospheric neutrino problem.
Because all three light neutrino masses are very small, the constrain from
the baryogenesis of the universe, which requires the primordial baryon
asymmetry is not washed out by the coexistence of B-L violation process for
neutrino mass generation and gauged B+L violation process, is satisfied. The
second choice is to have the mass of $\nu_{\tau}$ around 0.1eV, and mass of $%
\nu_{\mu}$ around $3\times 10^{-3} $eV and $\nu_e$ much lighter. In this
case three neutrino oscillation can possibly explain both solar and
atmospheric neutrino problems, but no candidate of hot dark matter is
provided. The third choice is with three neutrinos as heavy as about 1KeV,
5eV and $10^{-4}$eV. With this neutrino masses, $\nu_{\mu}$ may serve as the
candidate of hot dark matter and the oscillation between $\nu_{\mu}$ and $%
\nu_e$ can explain the LSND neutrino oscillation experimental data \cite{lsn}%
. However the KeV $\tau$ neutrino must decay fast enough in order not to
delay the beginning of matter dominated epoch of the universe too much. This
demands the lifetime $\tau (\nu_{\tau})\leq 2\times 10^{2}(\frac{1KeV}{%
m_{\nu_{\tau}}})^2 yr$ \cite{kim1}. In our model the dominant decay modes
for $\nu_{\tau}$ are $\nu_{\tau}\to\nu_{(\mu,e)}+(\mu,e)^{\pm}$. Its
lifetime is therefore estimated as $\tau(\nu_{\tau})\geq 10^{12}(\frac{KeV}{%
m_{\nu_{\tau}}})^5 yr$. We see that this constrain along rules out the third
choice.

Now we proceed to discuss the majoron model. In the majoron model the relic
density of $\nu_R$ is not only determined by the annihilation processes ${%
\bar\nu_R}\nu_R\to\bar{\nu_l}\nu_l, e^+e^-...$ but also by the process $%
\nu_R\nu_R\to\phi_R\phi_R$, here $\phi_R$ is the majoron associated with
spontaneous lepton number breaking. The coupling between $\phi_R$ and other
standard model particles is only induced by loop effect and proportional to
some power of Yukawa coupling, so it is negligible in considering of the
relic density of $\nu_R$. We estimate the cross section for the second
annihilation process $\nu_R\nu_R\to\phi_R\phi_R$ as 
\begin{equation}
<\sigma_Av>\sim\frac{h^4}{3\pi M_R^2}(\frac{P}{E})^2\simeq\frac{h^4T} {\pi
M_R^3} 
\end{equation}
in terms of the energy E and three momentum P of $\nu_R$ in the center of
mass frame, and T is the temperature of the universe. It is noticed that for
this process it is the p-wave dominated. The s-wave contribution is
forbidden as a result of momentum and CP conservation as well as the
statistics. This is roughly a weak interaction cross section if $h$ is
around order of one and $M_R$ at weak scale. Since the second process
dominates over the first annihilation process, it is the second annihilation
cross section determines the relic density of $\nu_R$ at present. To get a
feeling of the numbers, $<\sigma_A v>\sim 10^{-37}cm^2$ with $h\sim 0.1$ and 
$M_R\sim 100$ GeV. Since $\phi_R$ decouples from the standard model
particles at a high energy scale $\sim M_R/20$ larger than a few GeV,
majoron contributes to the effective number of light neutrino species $%
N_{\nu}$ less than 0.1 when primordial nucleosynthesis commences. Hence the
condition $N_{\nu}\le 3.3$ at the time of nucleosynthesis \cite{wal} is
satisfied. Other restriction from cosmology and astrophysics can also be
easily obeyed. The strongest one is due to the cooling of red giants. It
requires the coupling between electron pair and majoron is weaker than $%
10^{-11}$ \cite{mor}. In our model this coupling is safely smaller than this
number because this coupling is induced through one loop diagram and is
proportional to the square of electron mass.

The distinguished feature of the majoron model is that it offers new decay
channels for the heavier light neutrino. In our model $\nu _\tau $ can decay
to other two lighter neutrinos plus a majoron. We consider one interesting
situation here, mass of $\nu _\tau $ is about order of 10 KeV, $\nu _\mu $
is a few eV. In previous model without majoron this possibility is ruled
out. However due to the new decay channel to majoron the life time of $\nu
_\tau $ can be much shorter. The dominant decay channel is to $\nu _\mu $
plus a majoron. We estimate the life time of $\nu _\tau $ as 
\begin{equation}
\tau (\nu _\tau )\simeq 16\pi (\frac{m_{\nu _\tau }}{M_R})^{-4}m_{\nu _\tau
}^{-1}=10^3(\frac{10KeV}{m_{\nu _\tau }})^5(\frac{M_R}{100GeV})^4yr 
\end{equation}
Its dependence on the light neutrino mass is similar as that in the original
singlet majoron model, though the decay mechanism is different, in our model
nonvanishing contribution to this decay occurs at two loop level. To see
what kind role the $\tau $ neutrino can play in the cosmology, we need to be
more specific. We take $M_R=50$ GeV and $m_{\nu _\tau }=30$ KeV and find the
life time $\tau \simeq 0.2yr$. A neutrino with this mass and life time can
just be a late decaying particle which is required for the large scale
structure formation of the universe in the cold dark matter model \cite{kim1}%
. Nevertheless at the same time we have to require the $\nu _\mu $ to be
lighter than a few eV in order not to violate the same requirement. The mass
hierarchy between $\nu _\tau $ and $\nu _\mu $ is about one order of
magnitude larger than that expected from above mentioned seesaw relation $%
m_\tau ^2/m_\mu ^2$, though we think it is still reasonable. Here again we
see the advantage of the radiative seesaw model. Even the mass of $\nu _\tau 
$ is as small as 10 KeV, $M_R$ can be around 100 GeV duo to the loop factor.
So that $\nu _\tau $ can decay fast.

In conclusion, we discussed an new version of seesaw mechanism which is
realized radiatively and gave a concrete model to exhibit its interesting
new features. Generally speaking, in this mechanism the constrains from
cosmology and astrophysics are relaxed compared with that in the tree-level
seesaw model. we emphasize and focus on the weak scale seesaw in our model.
The most interesting application of this new mechanism is on cosmology and
neutrino physics. We pointed out that the lightest right-handed neutrino $%
\nu _R$ can be good candidate of cold dark matter. And at the same time
light neutrino may provide a hot dark matter or late decaying particle for
large scale structure formation, or offer solutions to other problems in
neutrino physics. Finally we would like to point out that if $\nu _R$ is the
dark matter of the universe, there are two possible ways to find out its
signal. The first is through high energy collider experiment. $\nu _R$ pair
can be produced through process like $e^{+}e^{-}\to {\bar \nu _R}\nu _R$.
Since $\nu _R$ is invisible and a Majorana particle, the best signal is to
search like sign charged lepton pair. Another way is to look for the
annihilation products $\mu ^{+}\mu ^{-}$ of dark matter $\nu _R$ pair in the
indirect dark matter search experiments.

This work is supported by the National Science Foundation of China (NSFC).


\begin{thebibliography}{99}
\bibitem{yan}  T. Yanagida, in Proceedings of Workshop on the Unified Theory
and the Baryon Number in the Universe, Tsukuba, Japan, 1979, edited by A.
Sawada and A. Sugamoto (KEK Report No. 79-188, Tsukuba, 1979); M. Gell-Mann,
P. Ramond, and R. Slansky, in Supergravity, Proceedings of the Workshop,
Stony Brook, New York, 1979, edited by D. Freedman and P. van Nieuwenhuizen
(North-Holland, Amsterdam, 1980).

\bibitem{gyu}  G. Gyuk and M. S. Turner, Nucl. Phys. B (Proc. Suppl.) 38, 13
(1995); M. Kawasaki {\it et al}, Nucl. Phys. B419, 105(1994); S. L. Glashow,
Phys. Lett. B 187, 367 (1987).

\bibitem{smi}  For a review on solar and atmospheric neutrino problems, see
A. Yu Smirnov, Talk given at 16th International Symposium on Lepton an
Photon Interactions, Ithjaca, NY, 10-15 Aug. 1993.

\bibitem{mdm}  Q. Shafi and F.W. Stecker, Phys. Rev. Lett. 53, 1292 (1984);
M. Davis, F. Summers, and D. Schlegel, Nature (London ) 359, 393 (1992); A.
van Dalen and R. K. Schaefer, Astrophys. J. 393, 33 (1992); A. A. Klypin 
{\it et al., ibid.} 416, 1 (1993).

\bibitem{dod}  S. Dodelson, G. Gyuk, and M. S. Turner, Phys. Rev. Lett. 72,
3754 (1994).

\bibitem{kim}  H. B. Kim and J. E. Kim; Nucl. Phys. B 433, 421 (1995).

\bibitem{cdm}  G. Blumenthal {\it et al}., Nature (London) 311, 517 (1984);
J. P. Ostriker, Ann. Rev. Astron. Astrophys., 31, 689 (1993).

\bibitem{kuz}  V. Kuzmin, V. Rubakov and M. E. Shaposhnikov, Phys. Lett.
B155, 16 (1985).

\bibitem{fuk}  M. Fukugita and T. Yanagida, Phys. Rev. D 42, 1295 (1990); J.
Harvey and M. S. Turner, {\it ibid.}, 3344 (1990).

\bibitem{pec}  For a review, see R. Peccei, in Proceeding of the XXVI
International Conference on High Energy Physics, Dallas, Texas, 1992; also
see J. Cline, K. Kainulainen and K. Olive, Phys. Rev. Lett. 71, 2372 (1993);
Phys. Rev. D 49, 6394 (1994).

\bibitem{tur}  N. Turok and J. Zadrozny, Nucl. Phys. B 369, 729 (1992).

\bibitem{bau} The concept of radiative seesaw mechanism was first proposed 
by Babu and Mathur some years ago, see  
K. S. Babu and V. S. Mathur, Phys. Rev. D38, 3550 (1988). In their paper 
they discussed a left-right symmetric 
model. However in this work we propose a different model to realize this 
mechanism and therefore some very interesting and general 
features, showing up in our model, for the radiative seesaw mechanism 
were not noticed by Babu and Mathur. 
\bibitem{pec1}  Y. Chikashige, R. Mohapatra and R. Peccei, Phys. Lett. B 98,
265 (1981); Phys. Rev. Lett. 45, 1926 (1980).

\bibitem{cal}  D. Caldwell {\it et al.}, Phys. Rev. Lett. 61 (1988) 510.

\bibitem{kol}  For a review, see The Early Universe, eds E. W. Kolb and M.
S. Turner (Addison-Wesley Publishing Company, 1990).

\bibitem{cha}  P. Chardonnet, P. Fayet and P. Salati, Nucl. Phys. B 394, 35
(1993).

\bibitem{lsn}  LSND Collaboration, Phys. Rev. Lett. 75, 2650 (1995); J. E.
Hill, ibid. 75, 2654 (1993).

\bibitem{kim1}  see equation (14) of reference 6.

\bibitem{wal}  T. Walker {\it et al.}, Astrophys. J. 376, 51 (1991).

\bibitem{mor}  M. Morgan and G. Miller, Phys. Lett. B 179, 379 (1986).
\end{thebibliography}
\end{document}